\documentstyle[twocolumn,aps,psfig]{revtex}
\tighten

\begin{document}
\draft

\title{
Spatial Inhomogeneities in Disordered d-Wave Superconductors:
Effect on Density of States and Superfluid Stiffness
}
\author{
Amit Ghosal, Mohit Randeria and Nandini Trivedi 
}
\address{Department of Theoretical Physics,
Tata Institute of Fundamental Research, Mumbai 400005, India \\
}
\address{
\begin{minipage}[t]{6.0in}
\begin{abstract}
We study a short coherence length d-wave superconductor with finite density of
unitary scatterers using the Bogoliubov-deGennes technique.
We find that the low-energy density of states is reduced,
the superfluid stiffness is significantly larger 
and off-diagonal long range order is more robust 
than the self-consistent T-matrix prediction.
These results are a consequence of the 
inhomogeneous pairing amplitude in the ground state and of the low-lying
excitations formed by hybridized impurity resonances.
These features, with their nontrivial spatial 
structure, cannot be adequately described within the 
conventional T-matrix approach.
\end{abstract}
\end{minipage}}
\date{\today}
\maketitle

\narrowtext
	Recently there has been a great deal of theoretical and 
experimental interest in the problem of disordered d-wave superconductors
(SC) given the d-wave symmetry of the high Tc cuprates.
A single unitary scatterer was predicted by Balatsky and coworkers
\cite{balatsky95} to lead to a 
low-energy resonance with a characteristic four-fold symmetric
wavefunction about the impurity site. This was recently observed 
in an STM study \cite{pan00}
of a Zn-doped cuprate. The $T$-matrix approximation used 
in Ref.~\onlinecite{balatsky95} is very accurate for the one-impurity problem,
and the order parameter suppression near the impurity, which it neglects,
does not lead to any qualitative changes. On the other hand, the problem
of a finite density of unitary scatterers is more subtle \cite{balatsky96}.
There is a large body of theoretical work using the self-consistent
$T$-matrix approximation leading to very interesting predictions
\cite{lee93,hirschfeld93,graf96,maki}.
However, the impurity averaging procedure used in this
method, and in other approaches \cite{senthil98} that go beyond it,
basically washes out the inhomogeneous structures that the disorder
potential gives rise to.

	We use the Bogoliubov-deGennes (BdG) approach
to study the spatial inhomogeneity induced by
unitary scatterers in a short coherence length ($\xi_0$) superconductor,
and to understand how these affect the low energy properties of the system 
such as the one-particle density of states (DOS) $N(\omega)$ and the 
superfluid stiffness $D_s$.
Our main results can be summarized as follows: \hfill\break
\noindent (1) 
The low energy DOS is considerably reduced relative to
the T-matrix result. The low lying excitations are found to be
generated by the interference of individual impurity resonances.
Such excitations, with their nontrivial spatial 
structure, cannot be adequately described within the 
T-matrix formulation. \hfill\break
\noindent (2) 
The superfluid stiffness $D_s$ is found to be significantly
larger than that obtained within the T-matrix analysis. The larger $D_s$
directly correlates with the lower DOS for ``normal'' excitations. 
\hfill\break
\noindent (3) 
We find that off-diagonal long range order (ODLRO)
and finite superfluid stiffness
survive to impurity concentrations much higher than the critical
concentration of the T-matrix approximation.  This relative insensitivity of
short coherence length d-wave superconductors to impurities is shown to
be closely tied to the inhomogeneity of the pairing amplitude on the
scale of $\xi_0$ in response to a random potential. 
In contrast, the T-matrix approach assumes
a uniform amplitude which then gets globally suppressed to zero at
a critical disorder.

	Several authors have previously used the BdG approach for
dirty d-wave systems. Tc reduction, superfluid density and localization of 
excitations was studied in Ref.~\onlinecite{franz96} 
and more recently the density of states has been studied in 
refs.~\onlinecite{flatte98,atkinson00}.
Our calculations differ from these in several aspects, some more important 
(e.g., working at fixed density rather than fixed chemical 
potential and inclusion of inhomogeneous Hartree-Fock shifts) 
than others (such as choice of Hamiltonian, parameters, 
and particle-hole asymmetry).
While our results are broadly consistent with those obtained previously,
what is new here is our emphasis on understanding the
BdG results for $N(\omega)$, $D_s$ and ODLRO in terms of
two different effects: 
(A) the inhomogeneity in the local pairing amplitude
which characterizes the disordered ground state, and (B) the spatial
structures characterizing the low-lying excitations 
in the disordered system. This provides a deeper insight into
our BdG results, and also highlights
the shortcomings of the $T$-matrix approach.


We model the 2D disordered d-wave SC by the Hamiltonian
${\cal H} = {\cal K} + {\cal H}_{\rm int} + {\cal H}_{\rm dis}$.
The kinetic energy 
${\cal K} = -t\sum_{<ij>,\alpha} (c_{i\alpha}^{\dag} c_{j\alpha} + h.c.)$
describes electrons, with spin $\alpha$ at site $i$ created by 
$c_{i\alpha}^{\dag}$, hopping between nearest-neighbors $<ij>$ on a 
square lattice.  The interaction term \cite{footnote1}
${\cal H}_{\rm int} = J\sum_{<ij>}\left({\bf S}_i \cdot {\bf S}_j
- n_i n_j /4 \right) + U \sum_{i} n_{i \uparrow} n_{i \downarrow}$
is chosen to lead to a $d$-wave SC ground state in the disorder-free
system. The spin operator 
$S^a_i = c_{i\alpha}^{\dag}\sigma^a_{\alpha\beta}c_{i\beta}$,
where the $\sigma^a$ are Pauli matrices, and the density
$n_{i\alpha}= c_{i\alpha}^{\dag}c_{i\alpha}$ with 
$n_i = n_{i \uparrow} + n_{i \downarrow}$.
Finally, ${\cal H}_{\rm dis} = \sum_i \left(V(i)-\mu \right) n_i$
where $\mu$ is the chemical potential and
the disorder potential $V(i)$ is an independent random
variable at each site which is either $+ V_0$, with a
probability $n_{\rm imp}$ (impurity concentration), or zero.
We believe that such a simple model is adequate to describe the
strongly-correlated cuprates at low temperatures
because their SC state has sharp quasiparticle excitations.

The BdG equations are given by:
\begin{equation}
\left(\matrix{\hat\xi & \hat\Delta \cr \hat\Delta^{*} & -\hat\xi^{*}} \right)
\left(\matrix{u_{n} \cr v_{n}} \right) = E_{n}
\left(\matrix{u_{n} \cr v_{n}} \right)
\label {eq:bdg}
\end{equation}
where
$\hat\xi u_{n}(j) = -\sum_{\delta}(t + W_j) u_{n}(j+\delta) + 
(V(j)-\tilde{\mu}_j)u_{n}(j)$ and 
$\hat\Delta u_{n}(j) = \sum_{\delta}\Delta(j+\delta;\delta) u_{n}(j+\delta)
$,
and similarly for $v_{n}(j)$. 
The pairing amplitude on a bond $(j;\delta)$, where 
$\delta = \pm{\hat{\bf x}}, \pm{\hat{\bf y}}$, is defined by
$\Delta(j;\delta) = -J\langle c_{j+\delta \downarrow}c_{j \uparrow}
+ c_{j \downarrow}c_{j+\delta \uparrow}\rangle/2$.
The inhomogeneous Hartee-Fock shifts are given by
$\tilde{\mu}_j = \mu - U \langle n_{j} \rangle/2
+ J \sum_{\delta}\langle n_{j+\delta} \rangle$
and $W_j = J \langle c_{j,-\alpha}^{\dag}c_{j+\delta, -\alpha} \rangle$

We numerically solve for the BdG eigenvalues $E_n \ge 0$ and eigenvectors 
$\left(u_{n},v_{n}\right)$ on a lattice of $N$ sites with periodic boundary 
conditions. We then calculate the pairing amplitude
$\Delta(j;\delta) = J\sum_n\left[u_n(j+\delta)v^*_n(j) 
+ u_n(j)v_n^*(j+\delta)\right]/2$ at $T=0$, the density
$\langle n_j \rangle = 2\sum_n |v_n(j)|^2$, and Fock shift
$W_j = J\sum_n v_n(j+\delta)v^*_n(j)$. 
These are fed back into the BdG equation, and the process iterated 
until self consistency \cite{footnote2} is achieved 
for {\it each} of the (local) variables defined on the sites and bonds of the 
lattice. The chemical potential $\mu$ is chosen to obtain a given
average density $\langle n \rangle = \sum_i \langle n_i \rangle/N$,
The d-wave pairing amplitude is given by
$\Delta(j) = \left[\Delta(j;+\hat{x})-\Delta(j;+\hat{y})
+ \Delta(j;-\hat{x}) - \Delta(j;-\hat{y}) \right]/4$.

We have studied the model for a range of parameters and lattice sizes. 
Here we focus on $J = U = 1.15$, in units of $t = 1$, with 
$\langle n \rangle=0.875$ (similar to the parameters used in
refs.~\onlinecite{franz96,wang95}) on systems of size up to $26\times 26$.
For these parameters, and $n_{\rm imp} = 0$, the DOS $N_0 \simeq 0.21$ and 
$\Delta_0 \simeq 0.077$ corresponding to a maximum gap of $0.31$.
For the impurity potential we choose $V_0 = 100$, 
close to the unitary limit. The results are averaged 
over 15 - 40 different realizations of the random potential. 


Let us first study the density of states (DOS)
$N(\omega) = {1 \over N}\sum_{n,i} \left[ |u_n(i)|^2\delta(\omega - E_n)
+ |v_n(i)|^2\delta(\omega + E_n) \right]$ (where we broaden the delta 
functions with a width comparable to average level spacing).
In Fig.\,1 we plot $N(\omega)$ for several impurity concentrations
on a small energy scale; for comparison,
the maximum energy gap in the disorder-free system is $0.31$
and the T-matrix self energy scale \cite{lee93} 
$\gamma = \sqrt{n_{\rm imp}\Delta/2N_0} \le 0.25$ 
for the parameters chosen; ($\Delta$ is the T-matrix gap).
In the T-matrix theory $N(\omega)$ is a constant for $\omega \le \gamma$,
while we find a sharp dip in the DOS close to the chemical potential,
consistent with ref.~\onlinecite{atkinson00}.
In fact, we found $N(0) = 0$ for {\it each}
impurity configuration at every concentration that we studied.
The scale of the sharp dip at finite $n_{\rm imp}$ was found to be
the same as the energy of an isolated impurity resonance.

It is very clear that
the low energy DOS in the BdG calculations is considerably smaller than 
that in the T-matrix approximation 
(even though we do not have the spectral resolution to quantify the 
asymptotic form of $N(\omega)$ as $\omega \rightarrow 0$). 
To highlight this, we compare
in Fig.~2 the finite $N(0)$ of the T-matrix analysis \cite{tmat}
with the BdG $\overline{N}(0)$, which
is the average of $N(\omega)$ over the (arbitrarily chosen) 
range $|\omega| \le 0.05 \ll \gamma$. 

\begin{figure}
\vskip-2mm
\hspace*{0mm}
\psfig{file=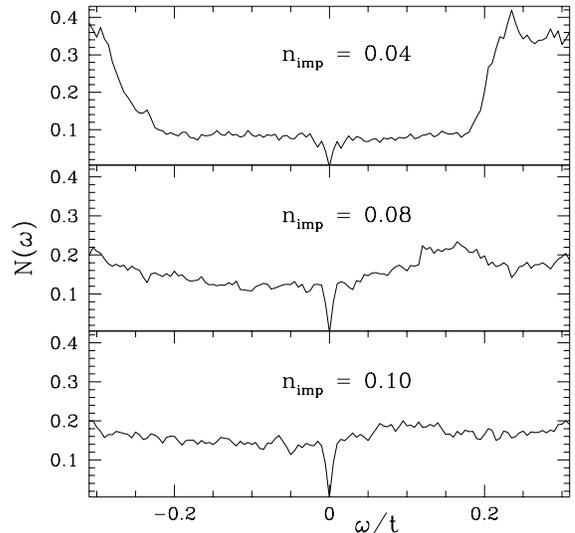,width=3.0in,angle=0}
\vskip0mm
\caption{
Density of states (DOS) on a $N=24\times24$ system, 
with $J=U=1.15t$ and $\langle n\rangle=0.875$, 
averaged over 40 disorder realizations at each $n_{\rm imp}$.
Note the sharp drop in the DOS near $\omega = 0$ on a scale
much smaller than the energy gap of $0.31t$ in the pure system.
}
\label{fig:dos}
\end{figure}

\begin{figure}
\vskip-10mm
\hspace*{0mm}
\psfig{file=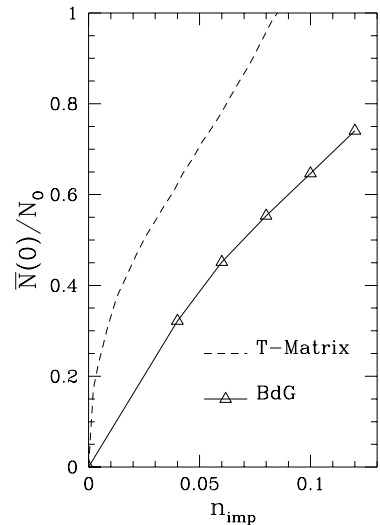,width=3.0in,angle=0}
\vskip0mm
\caption{
BdG density of states (DOS) $\overline{N}(0)$, 
defined as the average of $N(\omega)$ over the range $|\omega| \le 0.05$, 
is much smaller than the corresponding T-matrix result. The parameters are
the same as in Fig.~1 and the normalizing
factor is the pure system DOS $N_0 = 0.21$.
}
\label{fig:N0}
\end{figure}

To gain further insight into this difference between the T-matrix 
and BdG results, we study the wavefunctions of the low-lying excitations
for individual disorder realizations.
The probability density $|u_n(i)|^2 + |v_n(i)|^2$ corresponding to
the lowest energy states at various impurity concentrations are plotted
in right hand panels of Fig.~3. 
\begin{figure}
\vskip-2mm
\hspace*{0mm}
\psfig{file=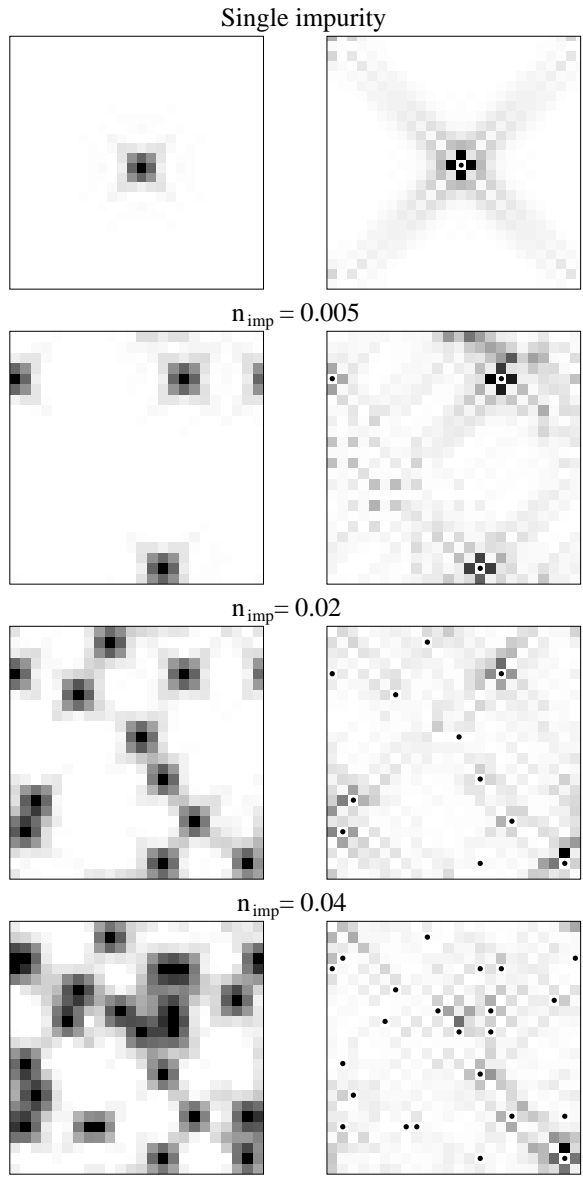,width=3.0in,angle=0}
\vskip2mm
\caption{
Left column: Evolution of the local pairing amplitude $\Delta(i)$ with impurity
concentration. Dark regions in the grey-scale plot indicate
suppressed pairing amplitude, and are correlated with the impurity locations. 
Parameters used are $J=U=1.15t$ and $\langle n\rangle=0.875$
on an $N=24\times 24$ system.
\hfill\break
\noindent Right column:
The corresponding probability density $|u_n(i)|^2 + |v_n(i)|^2$ for 
the lowest excited state ($n=1$) wavefunction. Higher probability is
indicated by a darker shade. Each impurity location is marked by a dot. 
}
\label{fig:panel}
\end{figure}

\noindent
The resonance for a single unitary impurity shows characteristic powerlaw
tails along diagonal directions \cite{balatsky95,footnote3}. From 
Fig.~3, and other low lying excitations not shown here, 
we see that for finite $n_{\rm imp}$ these wave functions are generated by the 
hybridization of individual impurity resonances. The effects
of constructive and destructive interference between the 
``diagonal tails'' of individual resonances are apparent.
The importance of such states was suggested in ref.~\onlinecite{balatsky96};
however, their analysis assumed that the resonance energies are randomly 
distributed over a scale $W \gg \Delta_0$, which is not the case in the
physical situation obtained here.

We emphasize that excitations with such non-trivial spatial structures
cannot be described by T-matrix theory, which treats the scattering of 
quasiparticles in a homogeneous (impurity averaged) medium off 
a single impurity in a self-consistent fashion. The resulting
constant $N(0)$ then arises from a constant broadening $\gamma$ 
(defined above) of states near the d-wave nodes.
In contrast, the low energy DOS in the BdG theory comes from
new states arising out of hybridization of impurity resonances.
 
We already see from Fig.~2 that at and beyond the critical concentration
of the T-matrix approach, $n^c_{\rm imp} \simeq 0.08$ for our
choice of parameters, the BdG DOS does {\it not} approach the non-disordered 
value $N_0$. This raises the questions: does SC persist beyond 
$n^c_{\rm imp}$, and if so, how? 
To address these issues we calculate the superfluid stiffness using the 
linear response result:
$D_{s}/\pi = \langle -k_{x} \rangle -
\Lambda_{xx}(q_{x}=0,q_y \rightarrow 0,\omega=0)$.
The diamagnetic term $\langle -k_x\rangle$,
is one-half (in 2D) the kinetic energy
$\langle-{\cal K}\rangle$, and the paramagnetic
term $\Lambda_{xx}$, is the long wavelength limit of the 
transverse current-current correlation averaged over 
disorder realizations.

\begin{figure}
\vskip-9mm
\hspace*{0mm}
\psfig{file=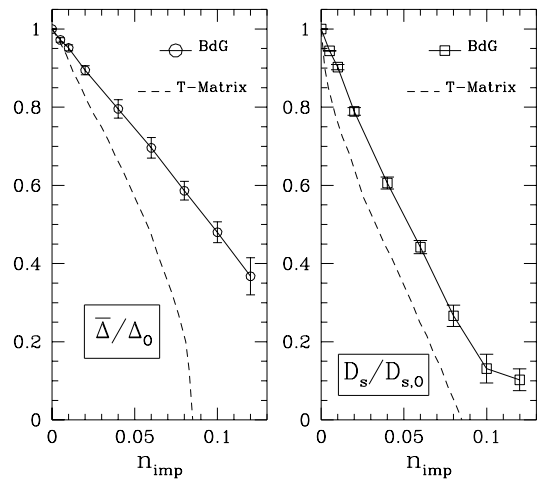,width=3.0in,angle=0}
\vskip0mm
\caption{
$T=0$ (a) off-diagonal long order parameter and (b)
superfluid stiffness, as a function of concentration of
unitary scatterers, obtained by the BdG method.
Note that d-wave superconductivity is much
more robust than the T-matrix prediction.
Parameters used are $J=U=1.15t$ and $\langle n\rangle=0.875$,
with $N_0 = 0.21$ and $D_{s,0} = 0.80$,
on an $N=24\times 24$ system, averaged over 15 disorder realizations.
}
\label{fig:delsfd}
\end{figure}

We see from Fig.~4b that the superfluid stiffness $D_s$ is much larger than  
the T-matrix result, consistent with Ref.~\onlinecite{franz96},
and does not vanish up to $n_{\rm imp} = 0.12$ which is $50\%$
larger than $n^c_{\rm imp}$ within the T-matrix approximation.
(We did not go to higher impurity concentrations because of the 
increase in computational time to reach self-consistency.)
In any case, we expect that once 
$D_s$ is sufficiently small, phase fluctuations neglected within
the BdG mean field approach will drive the transition to
the non-superconducting state \cite{ghosal98}; this is left for a 
future investigation. Here we wish to gain insight into {\it how} the 
system manages to exhibit $D_s > 0$, even when T-matrix theory 
predicts it to be non-superconducting.

One way to think about this is to correlate $D_s$ and $N(\omega)$.
A smaller DOS for low-lying excitations in the BdG approach implies fewer
``normal fluid'' excitations and hence a larger superfluid density
compared to the T-matrix approximation. 
A complementary approach, which we find very illuminating, relates the
$D_s$ to the inhomogeneous pairing amplitude $\Delta(i)$ 
in the disordered ground state, shown in the left panels of Fig.~3.
Notice that the d-wave pairing amplitude is suppressed in the vicinity of 
an impurity on the scale of the coherence length $\xi_0$ which is
3 to 4 lattice units. (In addition, a small extended s-wave component, not 
shown, also develops nearby). 
The regions of suppressed pairing amplitude give the appearance
of ``swiss cheese'' \cite{uemura} at finite $n_{\rm imp}$ in Fig.~3.

In the T-matrix approach the order parameter is forced to be
spatially uniform and it vanishes for $n_{\rm imp} \ge n^c_{\rm imp}$. 
However, by allowing the pairing amplitude to vary on the scale of
$\xi_0$, in response to the impurity potential, 
the BdG solution permits a non-vanishing order parameter 
$\overline{\Delta}$ which is larger than that obtained within T-matrix theory
for {\it all} $n_{\rm imp}$; see Fig.~4a.
($\overline{\Delta}$ is formally defined in terms of the long
distance behavior of the appropriate reduced two-particle density matrix).
We note that both $\overline{\Delta}/\Delta_0$ and $D_s/D_{s,0}$ are linear
functions of $n_{\rm imp}\xi_0^2$ for a substantial range of 
impurity concentration.

To qualitatively understand the superfluid stiffness $D_s$ consider 
applying an external phase twist to the inhomogeneous ground state. 
Despite the fact that at large $n_{\rm imp}$ there are large regions 
where the amplitude vanishes, there are still paths that permit phase 
information to be conveyed from one edge of the system to the other,
thus leading to a non-vanishing $D_s$.
Thus the spatial inhomogeneity of the pairing amplitude, which is
particularly important in short coherence length superconductors, is
crucial in understanding the relative insensitivity of the system to
unitary impurities, in that the order parameter and superfluid stiffness
are much larger than one might have guessed from the T-matrix approximation.
This lack of sensitivity of the high Tc cuprates to disorder has been 
seen in numerous experiments \cite{uemura}.

Despite the quantitative results on finite systems and their detailed
qualitative understanding, many questions remain open. 
The first one relates to Tc suppression. While it is easy to calculate
the ``mean field Tc'', a more reliable estimate should include the
effect of both phase fluctuations and quasiparticles.
Another important question is thermal transport \cite{taillefer}
in the SC state. Why does it not reflect the low energy structure
of the DOS and why is it consistent with the universal behavior 
predicted by T-matrix theory \cite{graf96,lee93}, when the superfluid
density \cite{uemura} shows deviations from it. 
A full understanding of the asymptotic DOS of the low-energy excitations, 
their localization properties and the study of SC state transport on a 
network of hybridized resonances are all topics for future research. 


{\bf Acknowledgements}: We would like to thank A. V. Balatsky, 
P. J. Hirschfeld, A. Paramekanti, S. H. Pan and G. P. Das for 
illuminating discussions.
M. R. was supported in part by the Department of Science and Technology
through the Swarnajayanti scheme.



\begin{references}
\bibitem{tmat}
The T-matrix results plotted in Figs.~2 and 4 are adapted from
ref.~\onlinecite{maki} for a particle-hole symmetric, continuum model
with the same values of $N_0$, $\Delta_0$ and $n_{\rm imp}$.
\bibitem{footnote3}
We will discuss elsewhere the dependence of the impurity resonance 
wavefunctions on model parameters and on the system size.
\end{references}
\end{document}